\documentstyle[aps,epsfig]{revtex}

\def\be{\begin{equation}}
\def\bea{\begin{eqnarray}}
\def\bma{\begin{mathletters}}
\def\ee{\end{equation}}
\def\eea{\end{eqnarray}}
\def\ema{\end{mathletters}}

\begin{document}

\title{Lower Bounds For Attainable Fidelities in Entanglement 
Purification}

\author{G. Giedke,$^{(1)}$ H. Briegel,$^{(1,2)}$ J. I. Cirac,$^{(1)}$ and P. Zoller$^{(1)}$}

\address{(1) Institut f\"ur Theoretische Physik, Universit\"at Innsbruck,
Technikerstr. 25, A--6020 Innsbruck, AUSTRIA\\
(2) Departamento de Fisica Aplicada, Universidad de Castilla--La Mancha,
13071 Ciudad Real, SPAIN}

\maketitle

\begin{abstract}
We derive lower bounds for the attainable fidelity of standard entanglement 
purification protocols when local operations and measurements are
subjected to errors. We introduce an error parameter which measures the
distance between the ideal completely positive map describing a
purification step and the one in the presence of errors. We derive 
non--linear maps for a lower bound of the fidelity at each purification
step in terms of this parameter.
\end{abstract}

\pacs{PACS number(s):}

\date{\today}

\widetext

\section{Introduction}

Entanglement purification \cite{Be96,Be96b,Gi96} is one of the most
important tools in the theory of Quantum Information and, in particular,
in Quantum Communication. It allows, in principle, to create maximally
entangled states of particles at different locations, even if the
channel that connects those locations is noisy \cite{Sc96}. These
entangled particles can be then used for faithful teleportation
\cite{Be93} or secure quantum cryptography \cite{Ek91,De96}.

The basic idea in entanglement purification is to ``distill'' a few $N'$
pairs of particles (qubits, for example) in highly entangled states out
of $N\ge N'$ pairs in a mixed state with lower fidelity of the
entanglement (or, in short, fidelity) using local operations and
measurements. This fidelity is defined as the maximum overlap of the
density operator of a pair of qubits with a maximal entangled state. If
the initial pairs are in a non--separable state \cite{Pe96,Ho96}, then
one can obtain asymptotically (in the limit $N\to \infty$) maximally
entangled states \cite{Ho96b} provided all local operations and
measurements are perfect \cite{Be96b,Ve98}. In practice, there will be
errors both in the local operations and measurements. The purpose of
this paper is to analyze this problem for the purification protocols
introduced in Refs.\ \cite{Be96,De96}. We are interested in analyzing
the conditions under which one can purify in the presence of errors, as
well as in the limitations of the purification protocols. In particular,
we find a non--linear map which relates a lower bound for the fidelity
at two consecutive steps of the purification protocol, which allows us
to derive lower bounds for the reachable fidelity. In order to analyze
this problem, we introduce a parameter $\delta$ which characterizes the
errors. It measures the distance between the ideal operations and
measurements and the ones in the presence of errors.

Quantum Communication in the presence of errors has been previously
considered by Knill and Laflamme \cite{Kn97} in a general context, and
by Van Enk {\it et al.} \cite{Va97} for a particular experimental setup
\cite{Ci97}. The work of Knill and Laflamme introduced ideas of
fault--tolerant quantum computation \cite{Sh96} to show that there
exists an accuracy threshold for storage of quantum information, which
also applies to the case of Quantum Communication. As shown by Bennett
{\it et al} \cite{Be96b} one can rephrase this result in terms of
entanglement purification with {\it one--way classical communication}.
In Ref.\ \cite{Br98}, entanglement purification together with a generic
error model is used to estimate the possibilities of quantum
communication over long distances using quantum repeaters. The employed
entanglement purification protocols explicitely utilize {\it two--way
classical communication}, which makes them much more efficient for
quantum communication. In the present paper we use purification
protocols which utilze {\it two--way classical communication}, and
therefore our lower bounds are much higher than those derived from the
theory of Knill and Laflamme \cite{Kn97}. On the other hand, we are
interested in a rigorous lower bound for the achievable fidelity for
arbitrary errors, and not in an estimation \cite{Br98}. The results and
methods developed here can be generalized to derive lower bounds for
other interesting problems in which local operations and measurements
are imperfect, such as quantum teleportation or quantum cryptography.

This paper is organized as follows: Section II contains a summary of the
main results of this paper, and is directed to the reader who is
interested neither in the technical details of the definitions of our
error parameter, nor in the derivations of the non--linear maps for the
lower bound of the fidelity. In Section III we introduce the error
parameter $\delta$ and derive some properties related to the fact that
it is a distance between completely positive linear maps. Finally, in
Section IV we derive the non--linear map for the fidelity of
entanglement in terms of this distance and sketch its dynamics.

\section{Summary of the main results and discussion}

In the standard scenario of entanglement purification \cite{Be96}, two
partners at different locations share $N$ pairs of qubits, each pair
being in a state described by a density operator $\rho$. A purification
procedure produces $N'\le N$ pairs in a state $\rho'$ ``closer'' to a
maximally entangled state $\psi_{\rm me}$ by only using local
operations, local measurements, and classical communication between the
partners. More specifically, if we define the fidelity of the
entanglement
\be
F(\rho)=\max_{\psi_{\rm me}} \langle\psi_{\rm me}|\rho|\psi_{\rm me}\rangle,
\ee
where the maximization is taken with respect to maximally entangled
states $\psi_{\rm me}$, then $F(\rho')>F(\rho)$. In the following we will call
$F(\rho)$ simply fidelity. 

It has been shown \cite{Ho96b} that if $\rho$ is non--separable (it
cannot be written as a convex combination of factorized density
operators \cite{Pe96,Ho96}) then there are purification procedures which
obtain $F(\rho')=1$ in the asymptotic limit $N\to\infty$. In particular,
if $F(\rho)>1/2$ one can reach this goal by using the {\it purification
procedure} devised by Bennett {\it et al} \cite{Be96} and improved by
Deutsch {\it et al} \cite{De96}. It consists of a concatenation of {\it
purification steps} involving two pairs of qubits, which give rise to a
single pair with higher fidelity. In all these procedures, one assumes
that the local operations and measurements are error free. In a real
situation, however, there will be errors due to the coupling to the
environment, imprecise apparatus, etc. Although small, they will limit
the maximum attainable fidelity and will dictate whether purification is
possible or not. 

In this section we first briefly review the purification protocol
introduced in Refs.\ \cite{Be96,De96}, and define the notation that we
will use later on. Then we consider the same procedure in the presence
of general errors, and characterize these errors in terms of a single
parameter, $\delta$, which basically expresses the departure of the
purification step in the presence of errors from the ideal one. Next, we
express the lowest possible fidelity (worst case) in each purification
step as a function of the lowest possible fidelity in the previouos step,
which leads to a non--linear map. We analyze this map and discuss the
conditions required for purification with imperfect means. The properties
of our definitions and the technical details are presented in the
following sections.

\subsection{Error free purification protocols}

In this subsection we review the two purification procedures introduced in
Refs.\ \cite{Be96,De96}. Subsequently we will refer to them as Scheme I and II, respectively. 
We characterize them in two different ways:
firstly, in terms of a completely positive linear map between the
initial density operator and the one after the measurement; secondly, in
terms of a non--linear map relating the diagonal matrix elements of the
density operator at each step in the Bell basis with the ones in the
previous step. In the next subsection we will generalize the first
characterization to the case of imperfect operations in order to
introduce the parameter describing the errors, and then we will
generalize the second characterization to find a lower bound for the
fidelity.

The purification protocols I and II both consist of a
sequence of \emph{steps} in which local operations are applied to two
pairs of qubits, followed by a measurement of one of the pairs which is
then discarded. Depending on the outcome of the measurement, the other
pair is discarded or not. In the latter case the fidelity
$F_1$ of the remaining pair is larger than that of the original ones. This
step is applied to the $N$ pairs obtaining $N_1\le N/2$ pairs of higher
fidelity $F_1$. Then it is applied to the resulting $N_1$ pairs
obtaining $N_2$ pairs of fidelity $F_2>F_1$. Continuing in this vein,
one can reach asymptotically $F_n\to 1$ when $n\to\infty$. 

Let us consider a single purification step. It starts out with two pairs
$1$ and $2$ in the state $\rho_{12}=\rho\otimes\rho$, applies the local
operations described by the superoperator $\cal U$ (unitary in the case 
of Scheme II) and then measures each
of the qubits of the second pair in the basis $\{|0\rangle,|1\rangle\}$.
We denote by $x$ the outcome of the measurement: $x=0$ if the qubits are
found in the state $|0\rangle_2\equiv|00\rangle_2$; $x=1$ if they are in
$|1\rangle_2\equiv|11\rangle_2$; $x=2$ if they are in
$|2\rangle_2\equiv|01\rangle_2$; and $x=3$ if they are in
$|3\rangle\equiv|10\rangle_2$ (the subscript $2$ denotes the second
pair). We denote by ${\cal P}_x$ ($x=0,\ldots,3$) the map defined as
follows
\be
\label{Pxideal}
{\cal P}_x (\rho_{12}) \equiv\, _2\!\langle x|{\cal U}(\rho_{12}) |x\rangle_2.
\ee
This map is linear and completely positive. The probability of
obtaining the outcome $x$ is $p_x(\rho_{12})={\rm tr}\left[ {\cal P}_x
(\rho_{12}) \right]$. If the outcome is $x=2,3$, then the first pair is
discarded and otherwise it is kept. In the latter case, the state of the
first pair will be 
\be
\label{rhop}
\rho_1' = \frac{{\cal P}_0 (\rho_{12})+{\cal P}_1 (\rho_{12})}
  {p_0(\rho_{12})+p_1(\rho_{12})}.
\ee
Thus, each step of the purification protocol is completely characterized
by the maps ${\cal P}_{0,1}$. (Note that ${\cal P}_x$ stand for different 
maps depending on whether we are discussing Scheme I or Scheme II.) 

On the other hand, if one is only interested in the fidelity at each
step, one can use a simpler characterization of each purification step
in terms of four real numbers. In the purification protocols I and II, the 
local operations characterized by $\cal U$ 
consist of a bilateral XOR gate and specific
single qubit rotations. In that case, the diagonal elements of the
density operator $\rho'$ in the Bell basis only depend on the diagonal
elements of the density operator $\rho$, and therefore each purification
step can be characterized by a non--linear map between these four
diagonal matrix elements. We denote by $A^i_n=\langle \phi^i|\rho_n|
\phi^i\rangle$ where $\rho_n$ is the density operator after the $n$--th
purification step and $|\phi^i\rangle$ are the elements of the Bell
basis ($i=0,1,2,3$), 
\begin{eqnarray*} 
|\phi^{1,4}\rangle &=& \frac{1}{\sqrt{2}}\left( |00\rangle \pm
 |11\rangle \right)\\ 
|\phi^{2,3}\rangle &=& \frac{1}{\sqrt{2}}\left( |01\rangle \pm
 |10\rangle \right).
\end{eqnarray*}

In particular, $A^0_n=F_n$, the entanglement
fidelity at each step. For Scheme II there is according to Ref.\ \cite{De96} a
simple non--linear map that relates $\vec A_{n+1}$ to $\vec A_{n}$, namely
\bea
\label{mapox}
A_{n+1}^i &=& \frac{\langle\phi^i|{\cal P}_0(\rho_n\otimes\rho_n)
 + {\cal P}_1(\rho_n\otimes\rho_n)|\phi^i\rangle}{{\rm tr}\left[
  {\cal P}_0(\rho_n\otimes\rho_n) + {\cal P}_1(\rho_n\otimes\rho_n)\right]}
  \nonumber\\
&=:& \frac{f^i(\vec{A}_n)}{g(\vec{A}_n)},
\eea
where 
\bma
\label{fig}
\bea
f^0(\vec A_n) &=& (A_n^0)^2+(A_n^1)^2,\\
f^1(\vec A_n) &=& 2A_n^2 A_n^3,\\
f^2(\vec A_n) &=& (A_n^2)^2+(A_n^3)^2,\\
f^3(\vec A_n) &=& 2A_n^0 A_n^1,\\
g(\vec A_n) &=& (A_n^0+A_n^1)^2+ (A_n^2+A_n^3)^2.
\eea
\ema
The map (\ref{mapox}) has a fixed point at $\vec A=(1,0,0,0)$, which is
reached if the initial state has $A^0_0=F>1/2$ \cite{Ma98}. This fact
expresses that in the absence of errors, one can use this purification
protocol to purify states with $F>1/2$ and reach a fidelity as close to
one as we please.

Scheme II \cite{Be96} is governed by a
similar map. The main difference is that at the end of each step the
resulting state is brought into Werner form, that is the three diagonal
elements $A^1,A^2,A^3$ are made equal to $(1-A^0)/3$. Therefore one can
concentrate on the first diagonal element, the fidelity $A^0$, only. The
fidelity after the $n$th purification step is then given by
\begin{equation}
\label{mapbe}
A^0_{n+1} = \frac{f^0(A^0_n,\frac{1-A^0_n}{3})}{g(A^0_n,\frac{1-A^0_n}{3})}.
\end{equation}
As (\ref{mapox}), this map has an attractive fixed point at $A^0=1$, and all
$A^0_0>1/2$ are attracted to it.

\subsection{Characterization of errors}

In practice, while performing the purification protocols errors will
occur, in the local operation as well as in the measurements. The
imperfections in the local operations can be accounted for by
substituting the action of the superoperator $\cal U$ in Eq.\
(\ref{Pxideal}) by the action of some other completely positive, trace
preserving linear map. The errors in the measurements will be related to
the following fact: in practice, the outcomes $x=0,1$ will be ultimately
attributed to the presence/absence of clicks in some kind of detectors.
Due to imperfections, the projection operators (or, more generally, POVMs)
corresponding to these clicks are not exactly the same as the
ideal ones (\ref{Pxideal}). Consequently, the probabilities of the
outcomes $x=0,1$ as well as the state remaining after the measurement
will differ from the ideal ones. In general, we can describe both
these erroneous operations and measurements in terms of a single completely
positive linear map $\tilde{\cal P}_x$ which does not necessarily preserve
the trace (we will use tildes in the case in which there are errors to
distinguish them for the error free case). That is, if the two pairs are
initially in the state $\rho_{12}=\rho\otimes\rho$, a purification step
yields the outcome $x$ with a probability $\tilde p_x(\rho_{12})= {\rm
tr}[\tilde{\cal P}_x(\rho_{12})]$. The state of the pair after the
measurement is 
\be
\label{rhoi}
\tilde \rho_1' = \frac{\tilde{\cal P}_0 (\rho_{12})+\tilde{\cal P}_1
  (\rho_{12})} 
  {\tilde p_0(\rho_{12})+\tilde p_1(\rho_{12})}.
\ee
Thus, as before, the maps $\tilde{\cal P}_{0,1}$ completely characterize
each purification step. 

We characterize the errors by a single parameter as follows:
\be
\label{delta}
\delta := \max_{x=0,1} d({\cal P}_x,\tilde{\cal P}_x),
\ee
where $d({\cal P},\tilde{\cal P})$ denotes a distance between ${\cal P}$
and $\tilde{\cal P}$. The explicit form of this distance is given in
Eq.\ (\ref{dis}) below. We emphasize that for a given set--up, one can (in
principle) perform local measurements to completely characterize $\tilde{\cal
P}_x$, and therefore obtain the value of $\delta$ experimentally
\cite{Po97,Ah98}. The error parameter $\delta$ has a clear physical meaning
since it measures the distance between the ideal process and the
erroneous one. We would like to remark here that due to the fact that 
there are measurements and postselection involved in the process,
we have to work with maps ${\cal P}_x$ that do not preserve the trace.
In Section III we discuss why it is adventageous to use those maps
instead of trace preserving maps.

Some remarks concerning the adopted description of errors are in order:
We envision $\cal P$ as the reduced dynamics of the two entangled pairs
coupled to some environment. As shown in \cite{Pe94} reduced dynamics
need in general not be completely positive (not even positive) on the
whole system space. In taking the imperfect system dynamics to be
completely positive we do (as discussed in \cite{Pe94}) essentially
assume that there is \emph{no initial entanglement} between the system
and any environment to which it might be coupled during gate operations.
There may be, however, initial entanglement of the system with another
environment that is not affected by the gate operations. As in the
error--free purification schemes \cite{Be96,De96} we also assume the two
pairs that participate in a purification step to be disentangled from
each other.

\subsection{Purification with imperfect means}

Once we have defined a parameter that characterizes the errors at each
purification step, we can analyze the whole purification procedure
\cite{Be96,De96} in the non ideal case. In order to do that, we define
$\tilde A^i_n=\langle \phi^i| \tilde\rho_n| \phi^i\rangle$ where
$\tilde\rho_n$ is the density operator after the $n$--th purification
step. We are particularly interested in the fidelity at each
step $\tilde A^0_n=\tilde F_n$. In Section IV we show that 
for suitable initial conditions $\vec A_0$ and error parameter $\delta$
\be
\label{ineq}
\tilde A^0_n \ge a_n, \quad \tilde A^1_n \le b_n, \quad (n=1,2,\ldots)
\ee
where 
\bma
\label{nlmap}
\bea
\label{nlmapa}
a_{n+1} &=& \frac{a_n^2+b_n^2-2\delta}{(a_n+b_n)^2+(1-a_n-b_n)^2+2\delta} \\
\label{nlmapb}
b_{n+1} &=& \frac{(1-a_n)^2/2+2\delta}{a_n^2+(1-a_n)^2-2\delta} 
\eea
\ema 
and $a_0=\tilde A^0_0$, $b_0=\tilde A^1_0$. For Scheme I only the 
fidelity $A^0_n$ and therefore the bound
(\ref{nlmapa}) with $b_n$ replaced by $(1-a_n)/3$ is relevant.

Equations (\ref{nlmap}) define a non--linear map that can be iterated
to yield a lower bound for the attainable fidelity $\tilde F_\infty \ge
a_\infty$ which depends on the value of $\delta$. In the following we
will analyze the map (\ref{nlmap}). 

Let us first concentrate on the fixed points $(a_f,b_f)$ of this map, 
and consider in particular Scheme II. In
Fig.\ 1 (solid line) we have plotted $a_f$ as a function of the error
parameter $\delta$. For small values of $\delta\alt 0.01$ there are
three fixed points. The ones with largest and the smallest value of
$a_f$ are attractive, whereas the intermediate one is a saddle point
attractive in one direction and repulsive in the others. For larger
values of $\delta$, only the smallest one survives. This means that for
the appropriate initial values of $a_0$ and $b_0$ if $\delta \alt 0.01$
one increases the fidelity using the purification protocol II to a value
larger than the one given by the right wing of the appropriate curve of
Fig.\ 1. For example, for $\delta\simeq 0.005$ one can obtain a fidelity
$F>0.95$. 

Now, let us analyze for which initial conditions $(a_0,b_0)$ the map
converges to the fixed point with the largest $a_f$, i.e., for which
purification is possible. In Fig.\ 2 we have plotted in the $(a,b)$
parameter space the curve (separatrix) between the stable regions for
several values of $\delta$ ($\delta_k=0.002k$, $k=0,1,\ldots,5$). For
any initial value $(a_0,b_0)$ lying to the right of each curve, the map
will converge to the corresponding fixed point (asterisks in the plot).
For $\delta=0.006$ ($k=3$ in the plot), for example, one can purify from
values of $a_0\agt 0.69$ up to values of $F\ge a_f\alt 0.94$; for
$\delta=0.002$, one can reach $F\alt 0.98$ starting from $a_0\alt 0.61$.
The results show that the error threshold for purification is much less
restrictive than the one for quantum computation \cite{Kn97}. 

\section{Distance between two positive maps}

We denote by $H$ a finite dimensional complex Hilbert space and by
$L(H)$ the complex Banach space of linear operators $A:H\to H$ with the
trace norm $||A||={\rm tr}(|A^\dagger A|^{1/2})\equiv {\rm tr}(|A|)$ (as
usual, $|A|\equiv|A^\dagger A|^{1/2}$). We denote by $C(H)\subset L(H)$
the convex set of positive linear operators $\rho$ acting on $H$ with
$||\rho||\le 1$, and by $P(H,H')$ the set of completely positive linear
maps ${\cal P}:C(H)\to C(H')$ fulfilling
\be
\label{posit}
||{\cal P}(\rho)|| \le ||\rho||
\ee
For positive operators, the trace norm simply coincides with the trace,
and therefore Eq.\ (\ref{posit}) is equivalent to
\be
{\rm tr}\left[ {\cal P}(\rho) \right] \le {\rm tr}(\rho)\le 1.
\ee

Given two completely positive maps ${\cal P},\tilde{\cal P}\in P(H,H')$, 
we define their distance 
\be\label{dis}
d({\cal P},\tilde{\cal P}) = 
\max_{\rho\in C(H)} || {\cal P}(\rho) - \tilde{\cal P}(\rho)||.
\ee
It is straightforward to show that $d$ is indeed a distance by
using the fact that the trace norm is a norm. 

With this definition, we can characterize the errors by using the
parameter $\delta$ as defined in (\ref{delta}). The motivation for this
definition with respect to other possible definitions is that it easily
gives lower bounds even for physical processes where there are
measurements and post selection (as it is in the case of entanglement
purification, cf. next section), i.e. when the map describing the
physical process is not trace preserving. On the other hand (although we
will not use this property here) it allows to easily bound the distance
between processes which are composed of several individual processes in
terms of the distances between the individual processes themselves (see
next subsection). 

One can define other distances between trace preserving maps: for
example, one can consider the map $\tilde{\cal P}'$ that transforms
$\rho_{12} \to \rho_1'$, where $\rho_1'$ is given in (\ref{rhoi}) in
terms of the linear maps $\tilde{\cal P}_{0,1}$. This new map, although
trace preserving, is nonlinear. If one defines distances between
$\tilde{\cal P}'$ and the corresponding (trace-preserving) ideal map
${\cal P}'$, problems related to the non--linearity arise: for example,
it can happen that while the distance $\delta$ between the linear maps
${\cal P}, \tilde{\cal P}$ is very small, the similarly defined distance
between the non--linear maps ${\cal P}', \tilde{\cal P}'$ is of the
order of 1, which makes the definition useless to derive bounds. The
reason is that low probability processes get ``magnified'' by the
normalization and then dominate the maximization used to define the
distance.

One can still define other error parameters to find sharper bounds to
the fidelity in entanglement purification. However, by increasing the
number of parameters one does not gain too much and the bounds become
more complicated to analyze. On the other hand, $d({\cal P}\otimes
1,\tilde{\cal P}\otimes 1) \ne d({\cal P},\tilde{\cal P})$ \cite{Ah98},
which would allow us to use $d$ in processes for which the system in
which we perform operations and measurements is entangled with another
system, without having to include the other system in the error
analysis. This may be useful, for example, in quantum computation where
operations are performed on single qubits that are entangled with many
other qubits. In that case, one can define other distances, as it is
done in Ref.\ \cite{Ah98}. In any case, in quantum communication if we
can bound the fidelity when the system is not entangled, we can
automatically derive a bound for the entanglement fidelity
\cite{Kn97,Sc96}.

\subsection{Properties of $d$}

In this subsection we derive some properties of the distance $d$ introduced
above. Given ${\cal P},\tilde{\cal P}\in P(H,H')$ we have:

\noindent
{\it (1)} We can restrict the maximization in (\ref{dis}) to one dimensional
projectors, i.e.
\be
\label{d2}
d({\cal P},\tilde{\cal P}) = 
\max_{\psi\in H,\\ |||\psi\rangle||=1} || {\cal P}(|\psi\rangle\langle\psi|) 
- \tilde{\cal P}(|\psi\rangle\langle\psi|)||.
\ee
{\bf Proof:} We just have to prove that the distance as given in
(\ref{d2}) is always larger or equal than the one given in (\ref{dis}),
since the converse is clearly true. For any $\rho\in C(H)$ we write
$\rho= \sum P_i |\phi_i\rangle\langle \phi_i|$ with $\sum_i P_i\le 1$
and $\psi_i$ normalized states of $H$. Using the linearity of ${\cal P}$
and $\tilde {\cal P}$ and that $||\sum_i P_i A_i||\le \max_i ||A_i||$, we
find that $||{\cal P}(\rho)-\tilde{\cal P}(\rho)|| \le \max_i ||
{\cal P}(|\phi_i\rangle\langle\phi_i|) - \tilde{\cal
P}(|\phi_i\rangle\langle\phi_i|)||$. Taking the maximum with respect to
$\rho$ in this inequality completes the proof. \hfill$\Box$\\

\noindent
{\it (2)} For all $\rho\in C(H)$ and $\phi\in H$ (normalized state) we have
\bma
\label{prop2}
\bea
\label{th1}
\langle\phi|{\cal P}(\rho)|\phi\rangle - d({\cal P},\tilde{\cal P}) \le 
& \langle\phi|\tilde{\cal P}(\rho)|\phi\rangle & 
\le \langle\phi|{\cal P}(\rho)|\phi\rangle + d({\cal P},\tilde{\cal P})\\
\label{th2}
{\rm tr}\left[{\cal P}(\rho)\right] - d({\cal P},\tilde{\cal P}) \le 
& {\rm tr}\left[\tilde{\cal P}(\rho)\right] & 
\le {\rm tr}\left[{\cal P}(\rho)\right] + d({\cal P},\tilde{\cal P})
\eea
\ema
{\bf Proof:} For (\ref{th1}) we use 
\be
|\langle\phi|{\cal P}(\rho)-\tilde{\cal P}(\rho)|\phi\rangle| \le 
||{\cal P}(\rho)-\tilde{\cal P}(\rho)|| \le d({\cal P},\tilde{\cal P}),
\ee
whereas for (\ref{th2}) we use
\be
\left|{\rm tr}\left[{\cal P}(\rho)-\tilde{\cal P}(\rho)\right]\right|
\le {\rm tr}\left[\left|{\cal P}(\rho)-\tilde{\cal P}(\rho)
\right|\right] = d({\cal P},\tilde{\cal P}).
\ee
\hfill$\Box$\\

Next, we give a property that allows one to bound the distance when
one applies sequential maps. This may be useful when one has a
concatenation of processes.

\noindent
{\it (3)} Given ${\cal P}\in P(H',H'')$ and ${\cal Q}\in
P(H,H')$, we define ${\cal P}\circ{\cal Q}\in P(H,H'')$ according to
$({\cal P}\circ{\cal Q}) (\rho) = {\cal P}[{\cal Q}(\rho)]$. Then, we
have
\be
d({\cal P}\circ{\cal Q},\tilde{\cal P}\circ\tilde{\cal Q})
\le d({\cal P},\tilde{\cal P}) + d({\cal Q},\tilde{\cal Q}).
\ee
{\bf Proof:} Using the properties of a distance, we have
\be
d({\cal P}\circ{\cal Q},\tilde{\cal P}\circ\tilde{\cal Q}) 
\le d({\cal P}\circ{\cal Q},{\cal P}\circ\tilde{\cal Q}) +
d({\cal P}\circ\tilde{\cal Q},\tilde{\cal P}\circ\tilde{\cal Q}).
\ee
On the one hand, we have
\bea
d({\cal P}\circ\tilde{\cal Q},\tilde{\cal P}\circ\tilde{\cal Q})
&=&  \max_{\rho\in C(H)} || {\cal P}[\tilde{\cal Q}(\rho)] 
   - \tilde{\cal P}[\tilde{\cal Q}(\rho)]|| \\
&\le& \max_{\rho'\in C(H')} || {\cal P}(\rho')] 
   - \tilde{\cal P}(\rho')|| = d({\cal P},\tilde{\cal P}),
\nonumber
\eea
where we have used (\ref{posit}) for $\tilde{\cal Q}$.
On the other hand, 
\bea
d({\cal P}\circ{\cal Q},{\cal P}\circ\tilde{\cal Q})  
&=&  \max_{\rho\in C(H)} || {\cal P}[{\cal Q}(\rho)] 
   - {\cal P}[\tilde{\cal Q}(\rho)]|| \\
&=&  \max_{\rho\in C(H)} || {\cal P}[{\cal Q}(\rho)-\tilde{\cal Q}(\rho)]||.
\nonumber
\eea
Now, since ${\cal Q}(\rho)-\tilde{\cal Q}(\rho)$ is self--adjoint, we can
substitute in this last equation its spectral decompostion 
\be
{\cal Q}(\rho)-\tilde{\cal Q}(\rho) = \sum_\phi |\phi\rangle\langle\phi|
\; \langle \phi|{\cal Q}(\rho)-\tilde{\cal Q}(\rho)|\phi\rangle
\ee
obtaining
\bea
d({\cal P}\circ{\cal Q},{\cal P}\circ\tilde{\cal Q})  
&=&  \max_{\rho\in C(H)} \sum_\phi 
  \left| \langle \phi|{\cal Q}(\rho)-\tilde{\cal Q}(\rho)|\phi\rangle \right|
  \;  || {\cal P}(|\phi\rangle\langle \phi|)|| \\
&\le&  \max_{\rho\in C(H)} \sum_\phi 
  \left| \langle \phi|{\cal Q}(\rho)-\tilde{\cal Q}(\rho)|\phi\rangle \right|\nonumber\\
&=&  \max_{\rho\in C(H)} || {\cal Q}(\rho)-\tilde{\cal Q}(\rho)||
  = d({\cal Q},\tilde{\cal Q}),
\eea
which completes the proof. \hfill$\Box$\\

\noindent {\it (4)} Finally, we show that the distance $d$ stems from
a norm, which may be 
useful to derive some other properties. First, let us enlarge the set $C(H)$
so that it becomes a Banach space. The simplest way is to define
$S(H)={\rm lin}_R \{C(H)\}$, that is, the set of operators that can be
written as a (finite) linear combination of positive operators with
real coefficients. The real Banach space $S(H)\subset L(H)$ is simply the space
of self--adjoint operators acting on $H$. In the same way, we can enlarge
the set $P(H,H')$. First, given a map ${\cal P}\in P(H,H')$ we define 
$\hat{\cal P}:S(H)\to S(H)$ by using the linearity of ${\cal P}$ [that is,
if $S(H)\ni A=\sum_i \lambda_i \rho_i$ with $\rho_i\in C(H)$, we define
${\cal P}(A)=\sum_i\lambda_i {\cal P}(\rho_i)$]. Then, we define
$Q(H,H')={\rm lin}_R \{ P(H,H')\}$ which is a real vector space. Using the
operator norm
\be
||{\cal P}||_{\rm op} = \max_{A\in S(H)\\||A||\le 1} ||{\cal P}(A)||,
\ee
it becomes a real Banach space. With this definitions we have

\be
\label{disp}
d({\cal P},\tilde{\cal P})
= || {\cal P}-\tilde{\cal P}||_{\rm op}.
\ee
\noindent
{\bf Proof:} We show that the distance given in (\ref{dis}) is smaller
or equal than the one defined in (\ref{dis}), since the converse is
obviously true since $C(H)\subset S(H)$. For any $A\in S(H)$ with
$||A||\le 1$ we can write $A=\sum_i\lambda_i|\phi\rangle\langle\phi|$,
where $\sum_i|\lambda_i|=1$. Now, arguing as in the proof of the property
(1), we obtain that $||{\cal P}(A)-\tilde{\cal P}(A)||\le \max_{\phi}
||{\cal P}(|\phi\rangle\langle\phi|)-\tilde{\cal
P}(|\phi\rangle\langle\phi|)||$. 
Taking the maximum over all possible $A\in S(H)$ we complete the
proof. \hfill $\Box$

The distance $d$ is not unrelated to other quantities used in
the literature to characterize erroneous operations. Typically, given
one of the other quantities, one can bound $d$ (and vice versa within
the respective domains of applicability). Specifically this is true
for the minimum fidelity, the error amplitude \cite{Kn97}, and the 
generic error model \cite{Br98}. The diamond norm introduced in
\cite{Ah98} is a generalization of the distance used here and
particularly useful to discuss operations on systems that are strongly
entangled with other systems.

\section{Non--linear map for entanglement purification}

In this section we derive the non--linear map (\ref{nlmap}) for the
bounds of the diagonal matrix elements in the Bell basis of the density
operator after each step of the purification process. Let us denote by
$\tilde A_n^i=\langle\phi^i|\rho_n|\phi^i\rangle, i=0..3$, where
$\rho_n$ is the density operator of a pair of qubits at the $n$--th
step. Analogous to (\ref{mapox}), we have 
\be
\tilde A_{n+1}^i = \frac{\langle\phi^i|\tilde{\cal
P}_0(\tilde\rho_n\otimes\tilde\rho_n) 
 + \tilde{\cal P}_1(\tilde\rho_n\otimes\tilde\rho_n)|\phi^i\rangle}
   {{\rm tr}\left[  \tilde{\cal P}_0(\tilde\rho_n\otimes\tilde\rho_n) + 
   \tilde{\cal P}_1(\tilde\rho_n\otimes\tilde\rho_n)\right]}
\ee
Using (\ref{prop2}) we have that 
\be\label{bounds}
\frac{f^i(\vec{\tilde{A}}_n) - 2\delta}{g(\vec{\tilde{A}}_n) + 2\delta}
  \le \tilde{A}_{n+1}^i \le 
  \frac{f^i(\vec{\tilde{A}}_n) + 2\delta}{g(\vec{\tilde{A}}_n) - 2\delta}.
\ee
where $f^i$ and $g$ are defined in (\ref{fig}). In the following
subsections we will discuss the two purification schemes separately in
detail.

\subsection{Scheme I}

As stated above for the scheme I we can use Eq.\ (\ref{mapbe}) instead
of $f^0$ and forget about the other three diagonal elements. This gives
\begin{equation}
\label{boundbe}
\tilde A^0_{n+1} \ge  \frac{(\tilde{A}_n^0)^2 +
(\frac{1-\tilde{A}_n^0}{3})^2 - 2\delta}{(\tilde{A}_n^0+\frac{1-\tilde A_n^0}{3})^2 +
(1-\tilde{A}_n^0+\frac{1-\tilde{A}_n^0}{3})^2 + 2 \delta}.
\end{equation}
Now we observe that the rhs of (\ref{boundbe}) is monotonically
increasing with $\tilde A^0_n$ for all $\tilde A^0_n\ge 1/8$. Therefore
replacing $\tilde A^0_n$ by $\frac{1}{8}\le a_n\le \tilde A^0_n$ in
(\ref{boundbe}) yields a lower bound for $\tilde A^0_{n+1}$. Since the
interval $[1/8,1]$ is mapped into itself by the lhs of (\ref{boundbe})
we arrive at the dynamical system defined by $a_0=A^0_0$ and
\begin{equation}
a_{n+1} =
\frac{a_n^2+(\frac{1-a_n}{3})^2-2\delta}{(a_n+\frac{1-a_n}{3})^2 +
(1-a_n-\frac{1-a_n}{3})^2+2\delta}.  
\end{equation}
For every $n$ the value of $a_n$ is a lower bound of the fidelity after $n$
purification steps. 

In the case $\delta=0$ the original map of Bennett \emph{et al} is
recovered. The three fixed points of that map at $a_l(\delta)\approx0.25,
a_i(\delta)\approx0.5$, and $a_u(\delta)\approx1$ survive even for nonzero
$\delta$ and are given by the roots of the cubic polynomial 
\[
x^3 - \frac{7}{4}x^2 + \left[\frac{7}{8}+\frac{9}{4}\delta\right]x -
\left[\frac{1}{8}-\frac{9}{4}\delta\right].  
\]
They are plotted as a function of $\delta$ in Fig.~1 (broken line).
For $\delta\ge0.008$ only the lower fixpoint survives. 

The upper and lower fixpoints are attractive, while the intermediate is
repulsive. Consequently even an imperfectly implemented Scheme I
allows to purify ensembles with initial fidelity $F_{in}>a_i(\delta)$ up
to a fidelity $F_{out}\ge a_u(\delta)$, provided that $\delta\le0.008$.

\subsection{Scheme II}

Scheme II converges faster than Scheme I and can tolerate somewhat
larger errors, but the analysis becomes significantly more complicated,
since all four diagonal elements of the density matrix come into play.
Using (\ref{bounds}) we have
\bma
\label{boundox}
\bea
\label{boundoxa}
\tilde A_{n+1}^0 &\ge& \frac{(\tilde{A}_n^0)^2 + (\tilde{A}_n^1)^2 - 2
\delta}{(\tilde{A}_n^0+\tilde A_n^1)^2 +
(\tilde{A}_n^2+\tilde{A}_n^3)^2 + 2 \delta}\\ 
\label{boundoxb}
\tilde A_{n+1}^1 &\le& \frac{2\tilde{A}_n^2 A_n^3 + 2 \delta}
{ (\tilde{A}_n^0+\tilde{A}_n^1)^2 + (\tilde{A}_n^2+\tilde{A}_n^3)^2 -
2 \delta}.  
\eea
\ema

To proceed the same way as in the previous subsection we need again a
monotonicity property of the right hand sides (rhs) of the Eqs.\
(\ref{boundox}) so that we can replace the values $\tilde A^i_n$ (which
are typically not known, since their exact value depends on the unkown
errors in $\tilde{\cal P}$) by lower or upper bounds, resp. \\ Using
$\sum_i \tilde A^i_n=1$ we can express the rhs of (\ref{boundoxa}) in
terms of $\tilde A^0_n, \tilde A^1_n$ only. It is straight forward to
check that the resulting expression is monotonically increasing in
$\tilde A^0_n$ and monotonically decreasing in $\tilde A^1_n$ for all
$(\tilde A^0_n,\tilde A^1_n)$ fulfilling 
\begin{equation}
\label{cov1} 
\tilde A^0_n\ge
\frac{1}{2}+\frac{3\delta}{1-2\delta} \,\,\mathrm{and}\,\, \tilde
A^1_n\le 0.5.  
\end{equation}
Thus provided that $\tilde A^0_n\ge a_n, \tilde A^1_n\le b_n$, and
$(a_n,b_n)$ fulfill the condition (\ref{cov1}) then $a_{n+1}$ as given
in Eq.\ (\ref{nlmapa}) is a lower bound for $\tilde A^0_{n+1}$.

It remains to justify Eq.\ (\ref{nlmapb}). Starting from
(\ref{boundoxb}) we can this time express the rhs only in terms of
$\alpha_n = \tilde A^2_n + \tilde A^3_n$ and $\beta_n = \tilde
A^2_n-\tilde A^3_n$ using the normalization condition: 
\[
\tilde A^1_{n+1} \le \frac{\frac{1}{2}(\alpha_n^2-\beta_n^2)
+2\delta}{\alpha_n^2 + (1-\alpha_n)^2 - 2\delta}.
\]
Now it is easy to check that the rhs of this inequality is monotonically
increasing in $\alpha_n$ (for fixed $\beta_n$) and takes (for fixed
$\alpha_n$) its maximum at $\beta_n = 0$, where we use the fact that
$\alpha_n\le 1- \tilde A^0_n$ and $\tilde A^0_n\ge0.5$. Since $\alpha_n =
\tilde A^2_n+\tilde A^3_n \le 1-\tilde A^0_n \le 1-a_n$ we arrive at Eq.\
(\ref{nlmapb}) by replacing $\beta_n\to 0$ and $\alpha_n\to 1-a_n$.

The discrete dynamical system defined by the map (\ref{nlmap}) has for
$0\le\delta\alt 0.01$ three fixpoints with $a$-coordinate around
$a_l\approx0.5, a_i\approx0.6, a_u\approx1$. Figure 1 (solid line) shows
them as a function of $\delta$. For $\delta>0.01$ only the lower
fixpoint survives. The exact $a$ values are given by the real roots of a
polynomial of seventh degree or equivalently by the intersections of the
curves $b_{n+1}(a)$ and
\begin{equation}
\label{bfix}
b_{fix}(a)=-a+\sqrt{a-(1+\frac{3}{2a-1})\delta},
\end{equation} 
the latter of which is defined by $a_{n+1}(a_n,b_{fix}(a_n))=a_n$. The
corresponding $b$-coordinates are $b_{n+1}(a_x)$, where $x=l,i,u$. 

As in the previous case the upper and lower fixpoints are attractive,
while the intermediate one is now a saddle point, attractive in one
direction and repulsive in the others. Now essentially the same argument
as in the previous subsection applies: points between intermediate and
upper fixed points are purified to a final fidelity $F_{out}\ge a_u$.
There are, however, two complications: First, the eventual fate of a
point $(a,b)$ depends on both $a$ and $b$. Second, we need to make sure
that the conditions (\ref{cov1}) are fulfilled in every step of the
iteration, otherwise it is no longer valid to interpret $(a_n,b_n)$ as
bounds of the actual values $(\tilde A^0_n, \tilde A^1_n)$. For both of
these complications we have been unable to find complete analytical
answers. Therefore we first give the numerical results before mentioning
partial analytical solutions.

Numerical calculations show that the physically meaningful set $\{(a,b)
: 0\le 1, 0\le b\le 1-a\}$ is divided in two parts by a curve passing
through the intermediate fixed point, the separatrix (see Fig.\~2).
Points to the right of that curve converge to the upper fixed point,
points to the left towards the lower one. Moreover the points to the
right do all satisfy the conditions (\ref{cov1}) and so do the orbits of
all these points. For all ensembles described by density matrices with
diagonal elements $A^0_0, A^1_0$ in that region, $a_n, b_n$ as defined
in (\ref{nlmap}) provide lower and upper bounds for the respective
fidelities after $n$ purification steps. For initial values to the left
of the separatrix our approach allows no statement. The case $\delta=0$
in Fig.~2 indicates, how many ``good'' points our
worst-case-consideration misses: as shown in \cite{Ma98} the exact
border of the set of purifyable points in the $(a,b)$-plane is given by
the straight line $a=0.5$.

For  a subset of the points to the right of the separatrix it is easy
to \emph{prove}
convergence: 
All the points $(a,b)$ fulfilling $a\ge a_i, b\le b_i$, 
and $a+b\le1$ converge to the upper fixed point $P_u$ (except for
$P_i$, of course).

{\bf Proof: } The proof proceeds in four steps. The main tool is the
monotonic dependence of $a_{n+1}, b_{n+1}$ on $a$ and $b$. (It is easily
checked by calculation, that the coordinates of the intermediate fixed
point satisfy the conditions (\ref{cov1}) for all $\delta$ so that
monotonicity holds.)

(i) Consider $(a,b)$ in the set enclosed by the two curves
$b_{n+1}(a)$  and $b_{fix}(a)$ (Ep.\ \ref{bfix}, cf. Fig.\ 3). For these points, we
have for all $n$ 
\[
a_{n+1}\ge a_n \,\,\mathrm{and}\,\, b_{n+1}\le b_n.
\]
Since $a_n$ and $b_n$ are bounded by the coordinates of the upper and
intermediate fixpoints, they form monotonical, bounded sequences and
converge therefore. Since $a_n$ increases and $b_n$ decreases,
they converge towards $(a_u,b_u)$.

(ii) Similarly it is seen that all points $(a\ge a_u,b\le b_u)$ do
converge to the fixed point ``from above''.

(iii) Now, consider a point $X = (a,b\le b_u)$ below the curve
$b_{n+1}(a)$. 

Let us call a point $(a,b)$ \emph{better} than $(a',b')$, if $a\ge a'$ and
$b\le b'$. Monotonicity implies that if  $(a,b)$ better than $(a',b')$
then this will also be true for the images of these points after one
iteration of the dynamical system. 

Now compare $X$ with $X'=(a'=a,b')$ between the curves
but with the same $a$ as $X$, and with $X''=(a''\ge a_u,b''=b)$.  
Clearly, $X$ is better than $X'$ but worse than $X''$. Since both $X'$
and $X''$ converge towards the upper fixpoint, so does $X$.

(iv) A similar argument applies, if we compare a point
$Y=(a,b>b_{fix}(a))$ with $Y'=(a'<a, b'=b)$ between the curves and
$Y''=(a''=a, b''\le b)$ below the curves: the primed points converge to
the upper fixpoint, and thus $(a,b)$ -- being better than $Y'$ and
worse than $Y''$ -- does so, too. This completes the proof.\hfill$\Box$

\section{Summary}
The entanglement purification protocols \cite{Be96,De96} in the
presence of errors in gate operations and measurements have been investigated. 
The errors are quantified by a single parameter derived from the trace norm.
We have shown that these protocols allow to increase the fidelity of the
entanglement even if implemented with imperfect quantum
gates and measurements, as long as the errors are below a 
threshold of the order 1\%.
We derived a non--linear map to calculated a lower bound for the
fidelity after $n$ purification steps. A polynomial is given, the root
of which gives a lower bound for the asymptotically attainable fidelity.

This work was supported in part by the
\"Osterreichischer Fonds zur F\"orderung der wissenschaftlichen
Forschung and by the European TMR network ERB-FMRX-CT96-0087.
G.G. thanks Wolfgang D\"ur for useful discussions.\\
Part of this work was completed during the 1998 Elsag-Bailey --
I.S.I. Foundation research meeting on quantum computation.

\begin{minipage}[t]{8cm}
\begin{figure}\label{Fpplot}
\epsfxsize=8.5cm 
\epsfysize=6.6cm 
\begin{center}
\epsffile{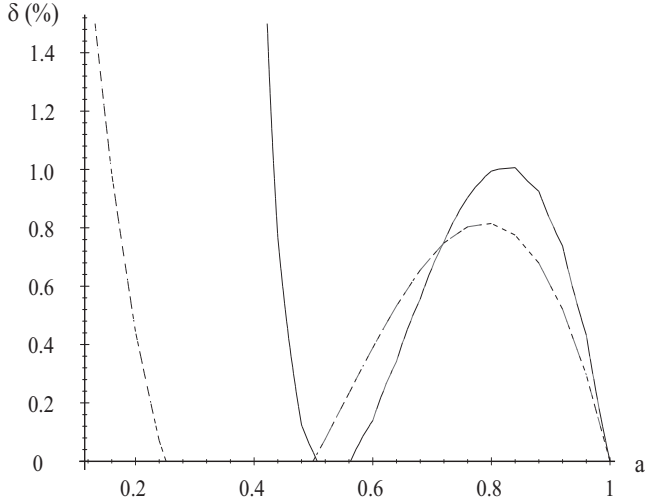}
\end{center}
\caption{The fixed points of the non--linear map: the intersections of
a horizontal line at $\delta$ with the plotted curve give the
$a$-coordinates of the fixed points for Scheme I (broken) and Scheme II (solid).} 
\end{figure}
\end{minipage}\hfill
\begin{minipage}[t]{8cm}
\begin{figure}\label{dynsys}
\epsfysize=6.6cm 
\epsfxsize=7.92cm 
\begin{center}
\epsffile{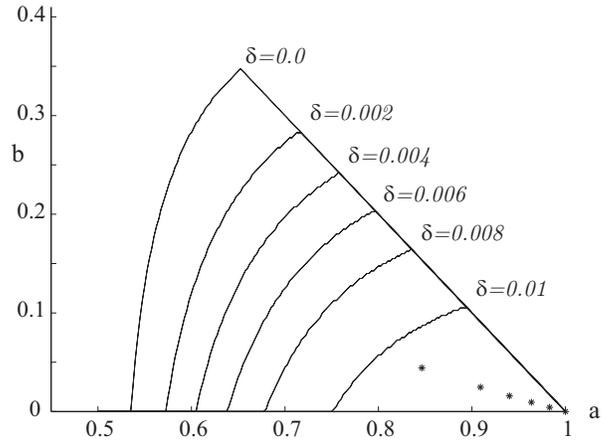} 
\end{center}
\caption{The solid lines show the border between the two
stable sets (the separatrix) for six values of
$\delta$. The asterisks show the corresponding ($\delta$ increasing from
right to left) upper fixed points.}
\end{figure}
\end{minipage}

\begin{minipage}[t]{8cm}
\begin{figure}\label{dynsyst}
\epsfxsize=8.5cm 
\epsfysize=6.6cm 
\begin{center}
\epsffile{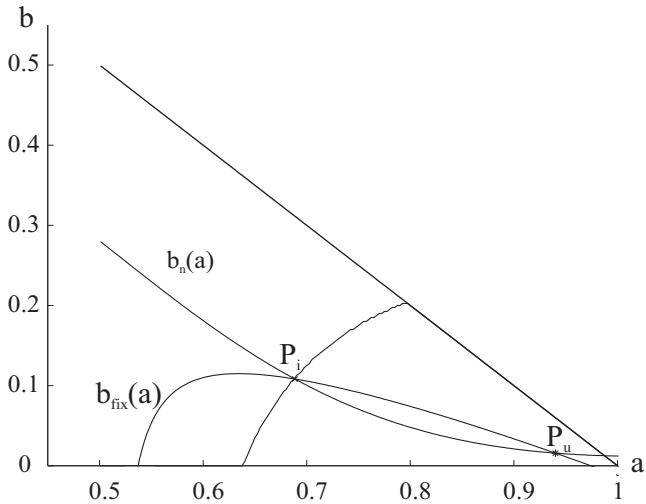}
\end{center}
\caption{For $\delta=0.006$ the curves $b_n$ (\ref{nlmapb}) and $b_{fix}$ (\ref{bfix}) 
are plotted. Their intersections are fixed points of the 
dynamical system.}
\end{figure}
\end{minipage}


\end{document}